# Thermally Oxidized MoS$_2$-Based Hybrids as Superior Electrodes for Supercapacitor and Photoelectrochemical Applications

Marzieh Rashidi, Foad Ghasemi*

Nanoscale Physics Device Lab (NPDL), Department of Physics, University of Kurdistan, Sanandaj, Iran, 66177-15175.

*Corresponding author: f.ghasemi@uok.ac.ir

**Abstract:**

High electronic transport and reasonable chemical stability of molybdenum disulfide ($MoS_2$) make it very suitable for electrochemical applications. However, its energy storage capacity is still low compared with other nanostructures. In this work, pristine and thermally oxidized $MoS_2$ ($O@MoS_2$) based hybrids are introduced by a simple method with enhanced capacitive performance thanks to the contribution of synergistic effects. Scanning electron microscopy (SEM), Transmission electron microscopy (TEM), X-ray diffraction (XRD), elemental mapping, UV-Visible, and Raman analyses are employed to investigate the morphological and crystalline structure of the introduced hybrids. In detail, the highest gravimetric capacitance of ~205.1 $Fg^{-1}$ is achieved for the $MoS_2$:$O@MoS_2$ hybrid with a mass ratio of 2:1 compared to pristine and other electrodes. This electrode is also accompanied by the longest discharging time and excellent cyclic stability of ~%113 after 2000 continuous charge-discharge cycles. In addition, photoelectrochemical testing of the introduced electrode leads to a ~63% increase in carrier photogeneration compared to $MoS_2$ due to the effective charge separation within the hybrid, which makes it suitable for water splitting and hydrogen production applications.

## 1. Introduction

Global population growth and the expanding energy use contribute to environmental pollution and climate change[1]. In recent decades, clean renewable energy technologies have received attention from researchers to remedy this issue. These attentions result in developing electronic devices such as supercapacitors[2], batteries[3], solar cells[4], and photoelectrochemical (PEC) cells[5] to generate/store clean energies. Among them, supercapacitors or electrochemical capacitors offer high power density, fast charge/discharge rate, and long cyclic life as energy storage systems[6, 7]. Supercapacitors are divided into electric double-layer capacitors (EDLC) and pseudocapacitors (PsC) based on their energy storage mechanism. In the first type, energy storage carries out by an accumulation of charges in the electrode/electrolyte interface[8], which is generally happened in materials with a large surface area such as porous carbon[9], carbon nanotubes[10] and graphene[11]. In the second type, charge storage occurs through faradic reactions or ion intercalations at the surface of electrode active materials[12] that are typically composed of transition metal oxides (TMOs) such as $NiO$[13], $V_2O_5$[14], $MnO_2$[15], $MoO_3$[16], $CuO$[17], and $RuO_2$[18].

Beside supercapacitors, PEC water splitting is a low-cost, sustainable, and environmentally friendly method to generate clean hydrogen fuel[19]. In this process, a semiconductor material with a visible light bandgap is needed to absorb the energy required to split water molecules[20]. PEC process mainly consists of four stages: light absorption, electron-hole generation, charge separation, and surface charge reaction that all together lead to the production of oxygen and hydrogen[21]. The efficiency of the PEC electrode is highly affected by light absorption, charge separation/transfer, and surface-active site density of the photoanode[22].

Until now, two-dimensional (2D) transition metal chalcogenide and oxide nanomaterials have been extensively investigated in energy storage/conversion fields due to their unique physicochemical properties[23]. Molybdenum disulfide ($MoS_2$), as a 2D structure, is widely used in energy devices thanks to its low cost, non-toxicity, ease of synthesis, high electrochemical activity, and desirable energy bandgap[6, 16, 20, 24, 25]. In general, the energy application of $MoS_2$ is divided into two categories: energy storage and energy conversion devices[26]. $MoS_2$ semiconductor is of great interest due to its excellent light absorption and high charge transport mobility[27]. It has an indirect bandgap of 1.2 eV in the bulk state and 1.8 eV in its monolayer state suitable for PEC-based devices[28]. Various $MoS_2$-based heterostructures have been so far introduced like rGO@$MoS_2$[29], $TiO_2$@$MoS_2$[30], ZnO@$MoS_2$[31], $MoS_2$@g-$C_3N_4$[32], and $MoO_3$@$MoS_2$[33] to improve the efficiency of the electrode by effective charge separation. In the case of energy storage applications, $MoS_2$ is also combined with other materials to increase its specific capacitance. Wang et al. synthesized a $FeS_2$@$MoS_2$ composite on molybdenum substrates that resulted in a large capacitance and long cyclic life due to good electrochemical stability of $MoS_2$ and high electrical conductivity of $FeS_2$[34]. Sarkar et al. obtained a volumetric capacitance of 343 $Fcm^{-3}$ for a $MoS_2$/rGO electrode at a scan rate of 10 $mVs^{-1}$ that is comparable with individual $MoS_2$ with a value of 212 $Fcm^{-3}$. This improvement originates from the ion intercalation behavior of $MoS_2$ and large active sites of rGO[35].

Molybdenum oxide phases, like molybdenum trioxide ($MoO_3$), are also widely used in energy storage systems. However, the poor electrical conductivity of $MoO_3$ greatly limits its applications in energy conversion/storage[36]. Nevertheless, its combination with other nanomaterials can address this issue via synergistic effects. Vattikuti et al. synthesized $MoO_3$ nanocrystals on $MoS_2$ nanosheets using a hydrothermal method with a larger surface area and improved electrical conductivity[37]. Liu et al. successfully synthesized a β-MoO$_3$@C hybrid, in which the presence of conductive carbon layers facilitates electron transfer reaction and ion intercalation. The prepared nanostructure shows excellent cyclic stability of 94% at a current density of 2 Ag$^{-1}$ after 50000 cycles[38]. However, most of these methods involve complex, time-consuming, and costly synthetic procedures that are also associated with limited electrical conductivity and an effective surface area. As a result, efforts to discover more efficient alternative methods are still underway. In this work, MoS$_2$ and its thermally oxidized structure (O@MoS$_2$) were integrated into supercapacitor electrodes using an easy and cost-effective method that is comparable with other reports that are accompanied with toxic precursors, time-consuming and complex processes[37, 39]. To prepare the hybrid, MoS$_2$ was mixed with O@MoS$_2$ in different mass proportions (4:1, 2:1, 1:1, 1:2, and 1:4) in N-Methyl-2-Pyrrolidone (NMP) solution, and then sonicated. Among all introduced electrodes, the capacitance and discharge time of the MoS$_2$:O@MoS$_2$ hybrid (2:1) increased by about 424% and 423% compared with pristine MoS$_2$, respectively. It also maintained long cycle stability with capacitance retention of ~113% after 2000 continuous charge-discharge cycles. In addition, the MoS$_2$:O@MoS$_2$ (2:1) hybrid demonstrated superior PEC performance than the MoS$_2$ due to the effective charge separation occurring between both materials.

## 2. Experimental

*2.1. Preparation of MoS$_2$ suspension*

NMP and MoS$_2$ powder (with initial flake size of <50 µm and 99% purity) were purchased from Sigma-Aldrich. 20 mg of MoS$_2$ powder was mixed with 10 ml of NMP solvent. The bulk powder was exfoliated in an ultra-sonication bath for 180 min. The sample was left for one day and the upper part of the solution was collected for analysis and fabrication of electrodes.

*2.2. Preparation of O@MoS$_2$ suspension*

Bulk $MoS_2$ and NMP was mixed and manually ground by mortar and pestle for 30 min. After being dried, the ground $MoS_2$ powder was placed in an electric furnace at different temperatures of (200, 300, 400, and 500 °C) for 60 min in air ambient to obtain its oxide phase. 20 mg of the obtained powder was mixed with 10 ml of NMP solvent and sonicated for 180 min. The upper part of the solution was collected after one day of standing.

*2.3. Preparation of $MoS_2$-$O@MoS_2$ suspension*

Different mass ratios of $MoS_2$ powder and $O@MoS_2$ were combined as follows: $MoS_2$:$O@MoS_2$ (4:1), $MoS_2$:$O@MoS_2$ (2:1), $MoS_2$:$O@MoS_2$ (1:1), $MoS_2$:$O@MoS_2$ (1:2), and $MoS_2$:$O@MoS_2$ (1:4). Then the obtained mixtures were added to 10 ml NMP solvent and sonicated for 180 min to prepare the corresponding suspensions.

*2.4. Fabrication of the working electrodes*

Stainless steel (SS) substrates were used as electrodes for supercapacitors. The obtained suspensions were stirred and drop cast on the SS substrates followed by air drying at room temperature. Copper wires were soldered to the backside of the substrates by silver paste. All openings (except for the surface of the substrate) were sealed with epoxy resin to isolate them from the solution. The surface area and mass of active material of all fabricated electrodes were measured.

*2.5. Electrochemical measurements*

The electrochemical performance of the electrodes was investigated using a three-electrode system at ambient temperature. The prepared working electrodes, Ag/AgCl reference electrode, and platinum counter electrode were used in 1 M $Na_2SO_4$ solution. Cyclic voltammetry (CV) tests were carried out under different scan rates of 5, 10, 20, 50, 100, and 200 $mVs^{-1}$ in a 0.0-0.9 V potential window. Galvanostatic charge-discharge (GCD) tests were measured in the range of 0.0-0.9 V vs

Ag/AgCl at different current densities. Electrochemical impedance spectroscopy (EIS) tests were done in a frequency range of 1000.00 kHz to 0.01 Hz in an open circuit potential of 0 V.

*2.6. Photoelectrochemical (PEC) measurements*

The PEC performance of the electrodes was evaluated based on the three-electrode photoelectrochemical system under exposure to a visible light source (200 W). Current-voltage characteristics of the electrodes were measured in dark and under light illumination at a scan rate of 50 mVs$^{-1}$ in 1 M $Na_2SO_4$ aqueous solution.

*2.7. Material characterization*

The X-ray diffraction (XRD) patterns of the samples were measured by a PANalytical X′pert Pro MPD diffractometer. The surface morphology of the electrodes was studied by a scanning electron microscope MIRA 3 LMU (Tescan Ltd) with Energy-Dispersive X-Ray Spectroscopy Mapping. Transmission Electron Microscope imaging was performed by Philips CM30. Raman spectra were measured by the Ocean Insight system (QEPRO-FL). The absorbance spectra of the suspensions were recorded using the UV-Vis-NIR Spectrometer (Varian).

**3. Results and discussion**

XRD analysis was performed to investigate the crystalline nature of the prepared samples. **Fig. 1** compares the XRD patterns of the pristine and thermally oxidized $MoS_2$ flakes at different temperatures of 200, 300, 400, and 500 °C for 60 min. The observed peaks at 14.5°, 29.1°, 32.7°, 33.6°, 39.6°, 49.8°, 58.4° correspond to (002), (004), (100), (101), (103), (105), (110) planes of 2H-$MoS_2$ crystal[40]. In the case of the thermally treated $MoS_2$ flakes, no sign of oxide formation observes at temperatures below 400 °C. In contrast, in the sample heated at 400 °C a weak peak is appeared at ~27.3° that is related to the (021) plane of $MoO_3$. In the 500 °C heated sample (named as O@$MoS_2$) two other peaks are also observed, the 23.3°, 25.6° peaks that refer to (110), and (040) planes of $MoO_3$ crystal, respectively[41, 42]. During this thermal process, the inner layers of the pristine MoS$_2$ flakes are expected to retain their structure while the outer layers are oxidized. However, in the thermally treated MoS$_2$ at 500 °C, both MoS$_2$ and MoO$_3$ phases are present. According to the inset of **Fig. 1**, the peak broadening of the O@MoS$_2$ in respect to the pristine MoS$_2$ attributes to an increase in strain or lattice deformation[43]. Moreover, the partial replacement of oxygen with sulfur is associated with the left-shift of the O@MoS$_2$ peaks due to the increase of the interlayer spacing [44].

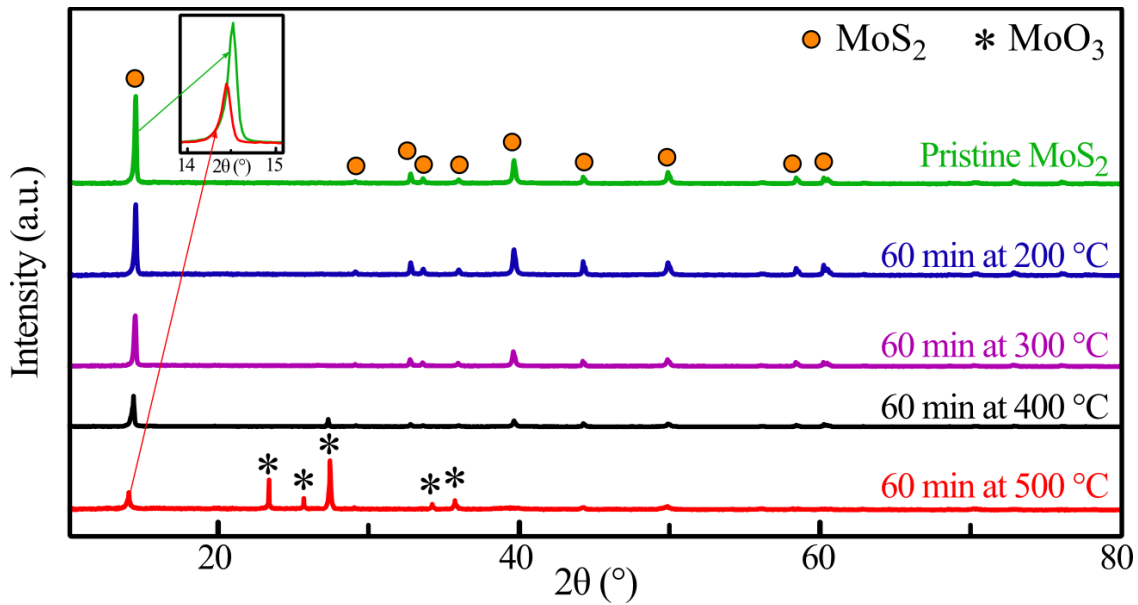

**Fig. 1.** XRD pattern of the pristine and thermally treated MoS$_2$ flakes at different temperatures.

Raman measurements were performed to further investigate the structure of the O@MoS$_2$ prepared at 500 °C. **Fig. 2** compares the Raman spectra of the pristine MoS$_2$ and prepared O@MoS$_2$ samples. In the case of the MoS$_2$, two peaks at ~397.7 and ~403.9 cm$^{-1}$ relate to the in-plane and out-of-plane oscillation modes, respectively[45]. In addition to these two peaks, five other peaks were found at 158.0, 279.9, 621.7, 821.5, and 993.0 cm$^{-1}$ for the O@MoS$_2$, which refer to the vibrational modes of MoO$_3$[46]. This result is also consistent with the XRD data that showed the presence of both structures in the O@MoS$_2$ flakes.

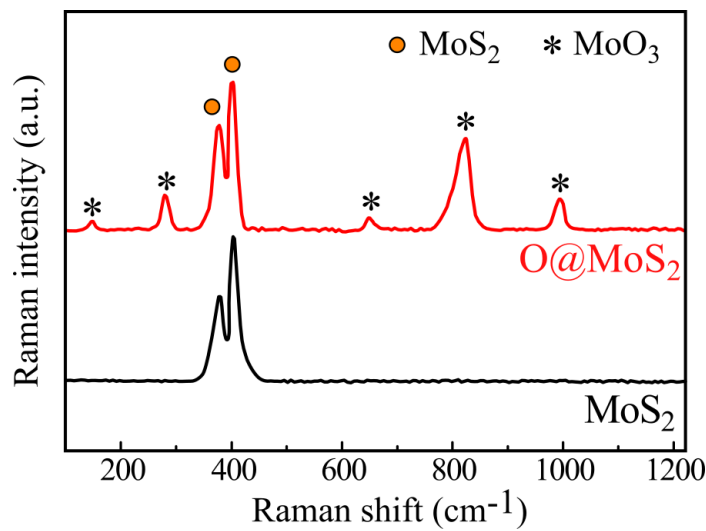

**Fig 2.** Raman spectra of the pristine and thermally treated MoS$_2$ flakes at 500 ºC for 60 min.

Electron microscope images of the prepared samples are obtained to characterize their morphological properties. SEM and TEM images of the MoS$_2$ flakes are shown in panels **(a)** and **(b)** of **Fig. 3**. Accordingly, few-layered structures of the MoS$_2$ can be seen in the TEM image where the selected area electron diffraction (SAED) pattern confirms its crystalline and hexagonal structure[47]. Interestingly, the SEM image (**Fig. 3(c)**) of the O@MoS$_2$ shows a spongy-like structure with pores originating from the sulfur removal in MoS$_2$. The porous structure of the O@MoS$_2$ is also observable in its TEM image presented in **Fig. 3(d)**. Its SAED pattern (Inset of **Fig. 3(d)**) indicates the polycrystalline nature of the O@MoS$_2$. The occurrence of porosity in the O@MoS$_2$ increases its effective surface area and provides more electrochemical active sites for energy storage. **Fig. 3(e-i)** demonstrate the SEM images of the samples with different mass ratios of 4:1, 2:1, 1:1, 1:2, and 1:4. As the mass ratio of each structure is dominant, the SEM image becomes more similar to that structure. For examples, the morphology of the 4:1 hybrid is much more similar to that of $MoS_2$ while the morphology of the 1:4 hybrid appears to be almost the same as a spongy-like structure of $O@MoS_2$.

Energy-dispersive X-ray-spectroscopy (EDS) was also used to determine the elemental composition and mapping of pristine and treated $MoS_2$ samples. **Fig. S1** shows the surface mapping of the pristine $MoS_2$ flakes with their dominant elements of O, S, and Mo. The surface mapping of $O@MoS_2$ with three main elements of O, S, and Mo is also presented in **Fig. S2**. Both mapping images show the uniform distribution of elements within the flakes. **Table S1** provides the atomic percentage of these three elements in all prepared samples. Accordingly, the Mo atomic percentage is measured to be 27.03 in the pristine $MoS_2$, which gradually decreases with the addition of $O@MoS_2$ in the composite samples, eventually reaching a minimum value of 17.34 in the $O@MoS_2$. In the same trend, the amount of S decreases from 51.42 in $MoS_2$ to 30.29 in $O@MoS_2$. By contrast, as expected, the content decrease of Mo and S elements is accompanied by an increase in O content. As a result, the oxygen atomic percentages are calculated to be 21.55 in the $MoS_2$ that increased to 52.37 in $O@MoS_2$. Thus, increasing the mass contribution of $O@MoS_2$ leads to an increase in the oxygen content of the final compounds.

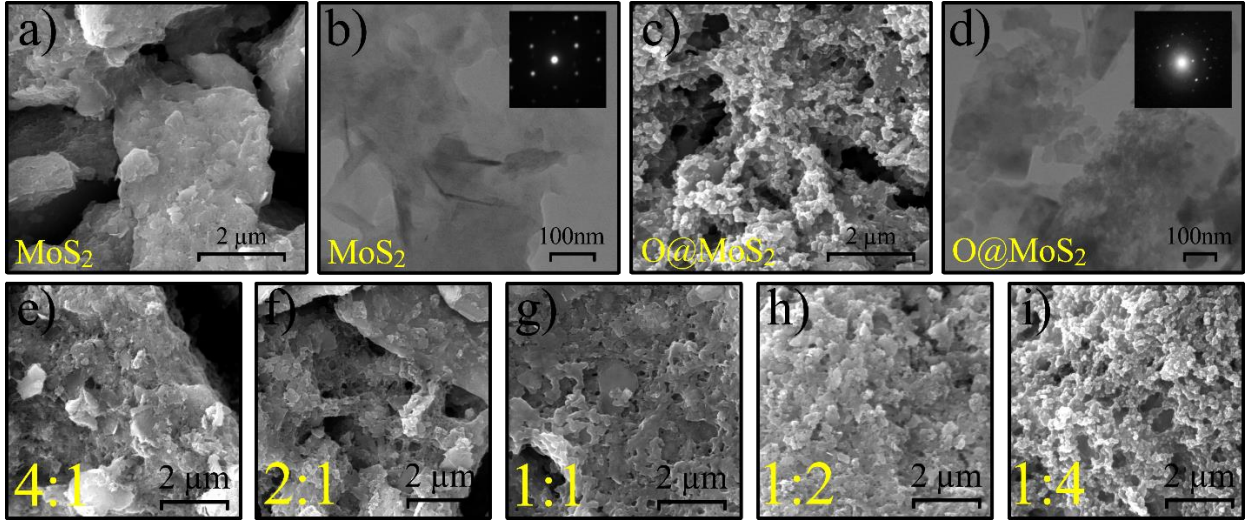

**Fig. 3.** SEM and TEM characterizations of the prepared samples. **(a)** SEM, and **(b)** TEM images of the pristine MoS$_2$ flakes. **(c)** SEM, and **(d)** TEM images of the O@MoS$_2$ flakes. SEM image of **(e)** 4:1, **(f)** 2:1, **(g)** 1:1, **(h)** 1:2, and **(i)** 1:4 composites.

After preparing the powder samples, they were sonicated in NMP for 180 min and the upper part of the solutions was collected. **Fig. S3** displays the photograph of the obtained suspensions. Furthermore, the UV-Visible spectra of the diluted solutions are presented in **Fig. 4(a)**. MoS$_2$ suspension exhibits two strong absorption peaks at 608 and 668 nm relating to excitonic states of B and A that originate from the direct transitions at the K point[48]. In the O@MoS$_2$ suspension, these two peaks almost disappear and absorption intensity strongly decreases in the visible region. Interestingly, the enhancement in the absorption intensity of the 2:1 solution could be due to the effective charge separation affected by the formation of the band offset between both materials[49]. Depending on the oxygen content, the absorption intensities of other suspensions distribute from MoS$_2$ to near O@MoS$_2$ spectra. The optical bandgap (E$_g$) can be calculated based on the Tauc formula through the absorption spectrum as follows[50]

$$(\alpha h\nu)^n = A(h\nu - E_g) \quad (1)$$

Where hν is the photon energy, α is the absorption coefficient, E$_g$ is the bandgap energy, and A is a constant. The value of n depends on the type of semiconductor optical transition in which n=2 is for an indirect allowed transition and n=1/2 for a direct allowed transition. The intersection of a linear fit with the hv-axis of the Tauc plot provides an estimate of the bandgap energy. Based on panels **(b)** and **(c)** of **Fig. 4**, $MoS_2$ and $O@MoS_2$ possess bandgap values of 1.99 and 2.89 eV, respectively. Based on the **Fig. 4(d to h)**, the different mass ratios of 1:1, 2:1, 4:1, 1:2 and 1:4 also modulates the energy bandgap to 2.34, 2.21, 2.05, 2.60, and 2.87 eV, respectively. Electronic interaction between the $MoS_2$ and $O@MoS_2$ results in the band-gap modulation in the composites due to good formation of the hybrid thanks to the efficient sonication process[51].

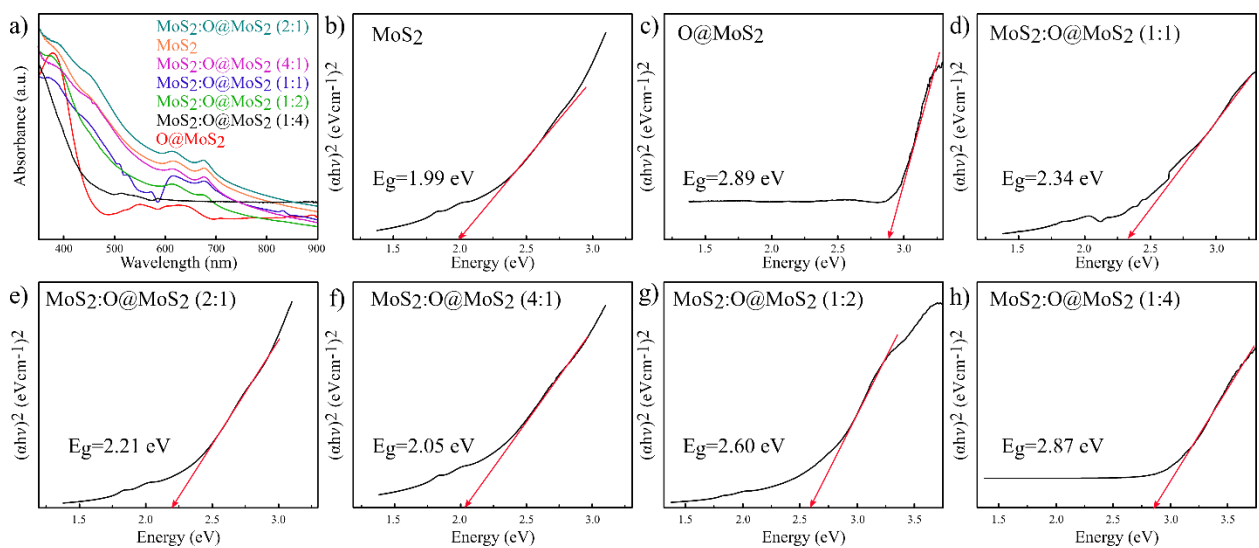

**Fig. 4. (a)** UV-Visible spectra of all introduced suspensions. Tauc plots of **(b)** $MoS_2$, **(c)** $O@MoS_2$, and $MoS_2:O@MoS_2$ with different mass ratios of **(d)** 1:1, **(e)** 2:1, **(f)** 4:1, **(g)** 1:2, and **(h)** 1:4.

The working electrodes were prepared by drop casting of the prepared solutions on the stainless-steel substrate according to the experimental section. **Fig. S4** provides the EDX spectrum of the $MoS_2$ film on a stainless-steel substrate containing the elements carbon, oxygen, molybdenum, sulfur, chromium and iron. **Table S2** shows the active material mass, active surface area, and active material/substrate weight ratio (mass loading) of the electrodes. Electrochemical testing of $MoS_2$, $O@MoS_2$, $MoS_2:O@MoS_2$ (1:1), $MoS_2:O@MoS_2$ (2:1), $MoS_2:O@MoS_2$ (4:1), $MoS_2:O@MoS_2$ (1:2) and $MoS_2:O@MoS_2$ (1:4) electrodes including cyclic voltammetry (CV), galvanostatic charge-discharge (GCD), and electrochemical impedance spectroscopy (EIS) analyses were performed in 1 M Na$_2$SO$_4$ electrolyte in a three-electrode system. **Fig. 5** displays the CV curves of the electrodes in a potential window of 0.0 to 0.9 V at different scan rates ranging from 5 to 200 mVs$^{-1}$. All CV curves have quasi-rectangular shapes showing the ideal capacitive performance[52]. It can be seen that by increasing scan rates, the current density is also increased since the ion diffusion rate increased with the sane rate[53]. Moreover, the area of CV curves becomes larger as the scan rate increases which does not mean more ion storage. Indeed, by increasing the scan rate, the ion-electrode interaction generally decreases, which also leads to a decrease in capacitance[54].

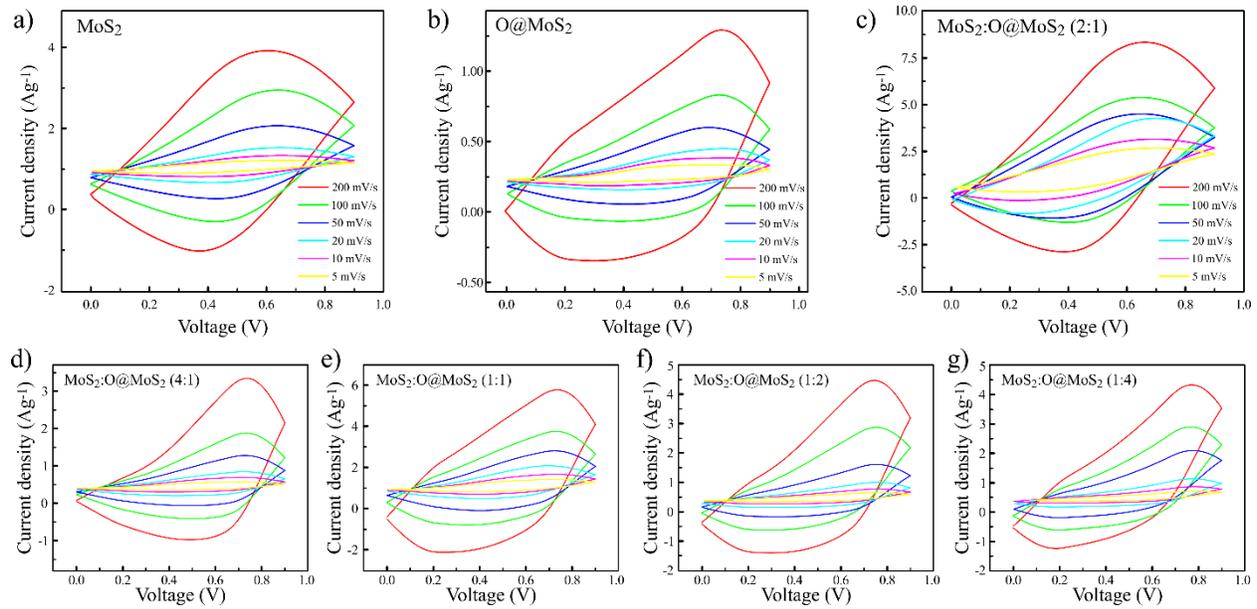

**Fig. 5.** CV curves of the **(a)** MoS$_2$, **(b)** O@MoS$_2$, and MoS$_2$:O@MoS$_2$ with different mass ratio of **(c)** 2:1, **(d)** 4:1, **(e)** 1:1, **(f)** 1:2, and **(g)** 1:4 at scan rates of 5 mVs$^{-1}$ to 200 mVs$^{-1}$.

Specific capacities of the electrodes are obtained from the following equations based on the CV curves[55]:

$$C_S(CV) = \frac{\int I(V)\, dV}{m\nu\Delta V} \qquad (2)$$

Where I (V) is the current, m mass of active material, ν scan rate, and ΔV potential window. The calculated unit of the specific capacity is expressed in terms of Fg$^{-1}$. The maximum capacity of

199.92 Fg$^{-1}$ is measured for MoS$_2$:O@MoS$_2$ (2:1) electrode at scan rate of 5 mVs$^{-1}$ while for MoS$_2$, O@MoS$_2$, MoS$_2$:O@MoS$_2$ with mass ratios of 1:1, 4:1, 1:2, and 1:4 the values of 30.56 Fg$^{-1}$, 23.82 Fg$^{-1}$, 48.09 Fg$^{-1}$, 22.62 Fg$^{-1}$, 27.40 Fg$^{-1}$, and 24.01 Fg$^{-1}$ are calculated, respectively. The spongy-like structure of the MoS$_2$:O@MoS$_2$ (2:1) electrode and the emergence of synergistic effects between both materials have improved its electrochemical performance compared to other electrodes. In contrast, it is expected that more contribution of O@MoS$_2$ in MoS$_2$:O@MoS$_2$ composite reduces the final conductivity of the electrode, which increases the charge transfer resistance and degrade its electrochemical performance.

**Fig. 5(a)** compares the CV curves of all electrodes at the same scan rate of 5 mVs$^{-1}$. The CV area of the MoS$_2$:O@MoS$_2$ (2:1) electrode is larger than other electrodes at this scan rate, especially in comparison to the pristine MoS$_2$ and O@MoS$_2$ electrodes, which suggests its better electrochemical activity and enhanced energy storage capacity. The gravimetric capacitances of all introduced electrodes are compared as a function of scan rate in **Fig. 5(b)**. In the case of the MoS$_2$:O@MoS$_2$ (2:1), the capacitance decreases from 199.92 Fg$^{-1}$ to 28.32 Fg$^{-1}$ by increasing the scan rate from 5 to 200 mVs$^{-1}$. For other electrodes, the capacitance similarly decreases with increasing scan rate due to the limited ion transfer at the surface of the electrode at higher scan rates[56]. In general, the O@MoS$_2$ surface provides more active sites for electrochemical reactions thanks to its high porosity. However, its oxide phase affects electrical conductivity and limits charge transport. Therefore, a trade-off is found between conductivity and the surface area of the final hybrids. According to the obtained results, it observes that in the 2:1 mass ratio of MoS$_2$ to O@MoS$_2$, the highest capacitance is obtained, which shows the optimal possible capacity in the combination of these two substances. The increase of the MoS$_2$ contribution leads to a decrease in the effective surface area of the hybrid. In contrast, the more contribution of O@MoS$_2$ results in the decrease of the final electrical conductivity of the hybrid. Moreover, the excess mass ratio of each material directly degrades the synergistic effect in the resulting component.

**Fig. 5(c)** shows a comparison of the GCD curves of the $MoS_2$, $O@MoS_2$, and hybrid electrodes in the voltage range of 0 to 900 mV at a current density of 0.5 $Ag^{-1}$. The GCD curves of the electrodes are approximately linear due to the contribution of both EDLC and PsC behaviors. The specific capacity is obtained from the following relation according to the charge-discharge curves[55]:

$$C_S = \frac{I \times t}{m \times \Delta V} \qquad (3)$$

Where I is the constant current, t discharge time, m mass of the active electrode material, $\Delta V$ potential window, and $C_s$ specific capacity in terms of $Fg^{-1}$. The $MoS_2:O@MoS_2$ (2:1) shows the highest capacitance value of 205.1 $Fg^{-1}$ compared to other electrodes at a constant current density of 0.5 $Ag^{-1}$. In addition, the $MoS_2:O@MoS_2$ (2:1) electrode has the longest discharge time of 369.2.9 s compared with $MoS_2$ (70.5 s), $O@MoS_2$ (41.9 s), 1:1 (106.3 s), 1:2 (54.4 s), 1:4 (46.0 s) and 4:1 (41.5 s) electrodes at 0.5 $Ag^{-1}$.

To further investigate the electrochemical properties of the electrodes, the EIS test is performed in a frequency range of 0.1 Hz to 1000 kHz at an applied potential of 0 V shown in **Fig. 5(d)**. In the Nyquist diagram, the intersection with the real axis of the impedance results in the equivalent series resistance ($R_s$) in the high-frequency range, and the diameter of the semicircle gives the charge transfer resistance ($R_{ct}$) in the middle-frequency range[57]. $R_s$ is a combination of the internal resistance of the electrode, the ionic resistance of the electrolyte, and the contact surface resistance at the electrode/electrolyte interface[58]. The $R_s$ values were measured to be 8.52 and 5.61 $\Omega$ for $MoS_2$, and $MoS_2:O@MoS_2$ (2:1) electrodes, respectively. The decrease in $R_s$ resistance of $MoS_2:O@MoS_2$ (2:1) compared with $MoS_2$ can be attributed to the increase of electrochemically active sites due to its spongy-like surface. Moreover, the $R_{ct}$ decreases from 0.72 $\Omega$ for $MoS_2$ to

0.58 Ω for MoS$_2$:O@MoS$_2$ (2:1) referring to the lower charge transfer resistance in the hybrid electrode.

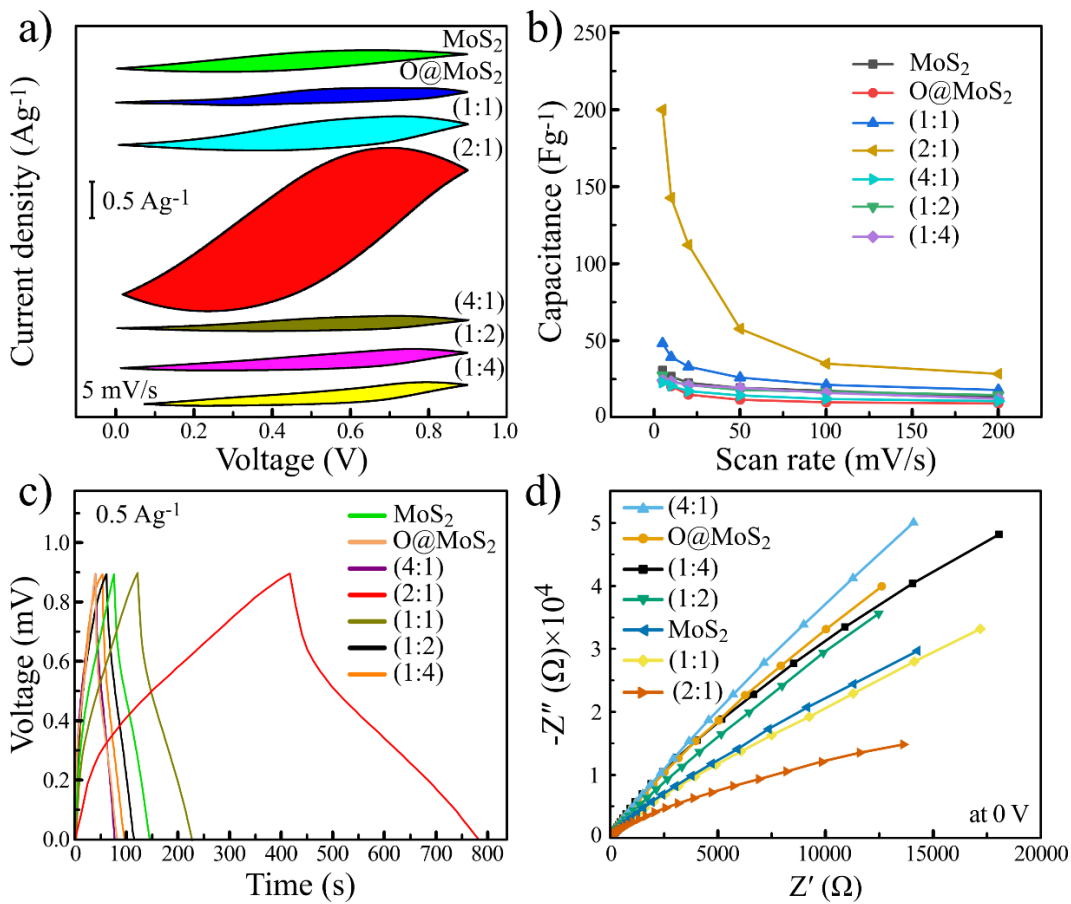

**Fig. 5. (a)** CV curves of electrodes at a 5 mVs$^{-1}$ scan rate, **(b)** Gravimetric capacitance of electrodes as a function of scan rate, **(c)** GCD curves at current density of 0.5 Ag$^{-1}$, and **(d)** Nyquist diagrams of electrodes.

**Fig. 6(a)** demonstrates the cycling stability of the MoS$_2$:O@MoS$_2$ (2:1) electrode after 2000 continuous charge-discharge cycles at a scan rate of 200 mVs$^{-1}$. The capacitance retention of ~113% is observed after 2000 cycles indicating an excellent long-term cycling performance of the introduced electrode. The capacitance retention value of more than 100% is due to the increase in available electrochemical active sites through self-activation of the active material during charge and discharge cycles[34]. The first and 2000th CV cycles of the electrode are presented in the inset of **Fig. 6(a)**. The larger area of the last CV curve well confirms the increase of the specific capacitance. To evaluate the energy densities of the $MoS_2:O@MoS_2$ (2:1), $MoS_2$, and $O@MoS_2$ electrodes, the Ragone plot is also calculated using GCD tests at a constant current density of 0.5, 1, 5, and 10 $Ag^{-1}$ in a two-electrode electrochemical cell. Energy (E) and power (P) densities are calculated through the following equations[59]:

$$E = \frac{C \times V^2}{2 \times 3600} \times 1000 \qquad (4)$$

$$P = \frac{E \times 3600}{t} \qquad (5)$$

Where C is the specific capacitance, V is the potential window, and t is the discharge time. According to **Fig. 6(b)**, the energy density of the $MoS_2:O@MoS_2$ (2:1) decreases from 15.38 to 4.75 $Whkg^{-1}$ by increasing power density from 224.98 to 4504.73 $Wkg^{-1}$. In the case of the $MoS_2$ and $O@MoS_2$, the highest energy densities are obtained at about 2.19 and 1,31 $Whkg^{-1}$ at power densities of 224.61 and 225.21 $Wkg^{-1}$, respectively. The higher energy density of the $MoS_2:O@MoS_2$ (2:1) hybrid with respect to the $MoS_2$ indicates its promising superior performance as an active compound for energy storage.

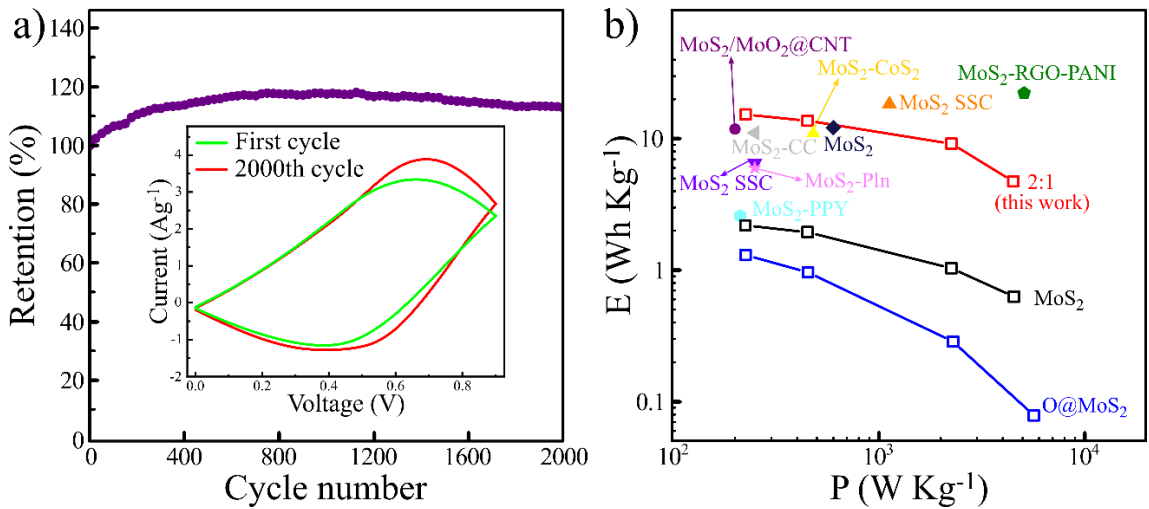

**Fig. 6. (a)** Cyclic stability of $MoS_2:O@MoS_2$ (2:1) electrode after 2000 cycles at 200 $mVs^{-1}$ scan rate. **Inset**: The first and last cycles after 2000 continuous charge and discharge cycles. **(b)** Ragone plot of the $MoS_2$, $O@MoS_2$ and $MoS_2:O@MoS_2$ (2:1) electrodes.

To evaluate the PEC performance of the electrodes, current-voltage measurements of the $MoS_2:O@MoS_2$ (2:1) and $MoS_2$ were carried out in dark and under visible light illumination. As shown in **Fig. 7(a)**, photocurrent density increases in both electrodes under light radiation. It is observed that the photocurrent density of the $MoS_2:O@MoS_2$ (2:1) hybrid shows a significant increase of 4.07 mAcm$^{-2}$ compared with $MoS_2$ (1.58 mAcm$^{-2}$) at 0.9 V voltage which can be due to the synergistic effect of the resulting hybrid, as well as efficient charge separation. As a result, the electron-hole recombination rate in the hybrid is less than that of $MoS_2$, which leads to more photogenerated carriers in the former[20]. A schematic illustration of the band structure and electron-hole separation in the $MoS_2:O@MoS_2$ (2:1) compound is presented in **Fig. 7(b)**. Since the energy bandgap of $MoS_2$ is smaller than that of the $O@MoS_2$, it plays the dominant role in absorbing visible light. After light irradiation, the excited electrons transfer from the valance band (VB) to the conduction band (CB) of the $MoS_2$, leaving holes in the VB. According to the Figure, the CB and VB of the $O@MoS_2$ are located below the $MoS_2$ bands. Hence, the photogenerated electrons are transferred to CB of the $O@MoS_2$, and the holes are injected into the VB of the $MoS_2$. The formation of the band offset within both structures prevents the recombination of electron-hole pairs. As a result, the efficient separation of photocarriers improves the current density of the hybrid that making it suitable for PEC applications.

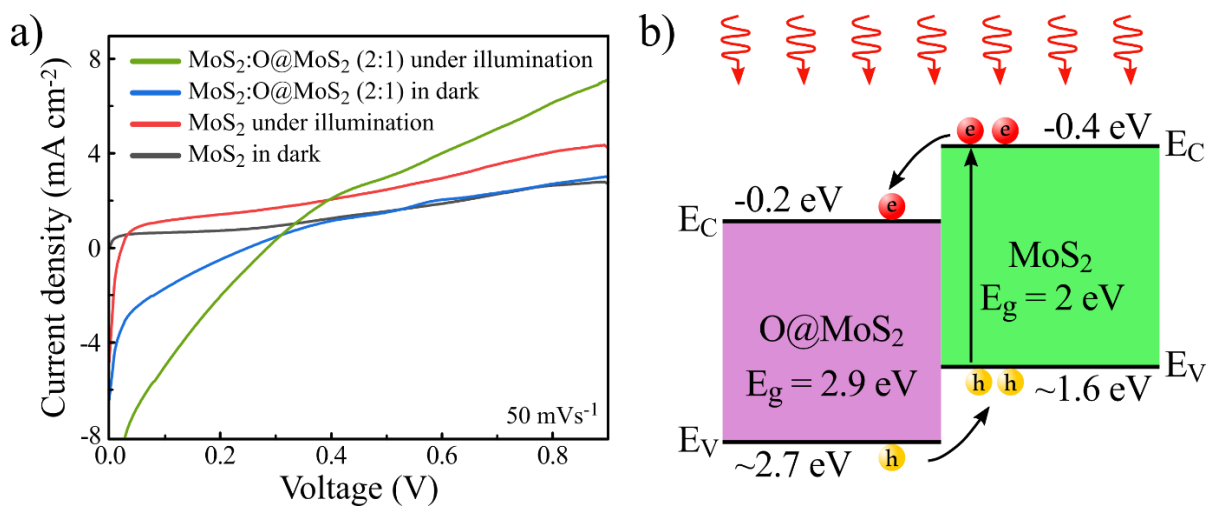

**Fig. 7. (a)** Linear sweep voltammetry measurements of the $MoS_2$:$O@MoS_2$ (2:1) and $MoS_2$ electrodes under visible light irradiation and in dark. **(b)** Depiction of charge separation process in the $MoS_2$:$O@MoS_2$ (2:1) hybrid.

**Table 1** compares the performance of the introduced hybrid with supercapacitors of a similar structure. According to them, the $MoS_2$:$O@MoS_2$ hybrid with an easy fabrication process shows a significant capacitance of ~200 Fg$^{-1}$ with high cyclic stability compared to other reports[60-64].

**Table 1.** Comparison of the supercapacitor performance of the $MoS_2$:$O@MoS_2$ electrode with $MoS_2$ and $MoO_3$ based electrodes in the literature.

| Active material | Electrolyte | Potential range | Scan rate | Gravimetric capacitance | ref |
|---|---|---|---|---|---|
| $MoS_2$ | 1 M $Li_2SO_4$ | -1.0 to 0.0 V | 5 mVs$^{-1}$ | 119.3 Fg$^{-1}$ | [65] |
| Pln/$MoS_2$ | 1 M $H_2SO_4$ | 0.0 to 1.0 V | 1 Ag$^{-1}$ | 173.0 Fg$^{-1}$ | [66] |
| $MoS_2$ SSC | 0.5 M TEABF$_4$ | 0.0 to 3.0 V | 25 mVs$^{-1}$ | 11.8 Fg$^{-1}$ | [67] |
| Ti@$MoS_2$ | 1 M KCl | -0.9 to -0.1 V | 10 mVs$^{-1}$ | 118.1 Fg$^{-1}$ | [68] |
| $MoO_3$@$MoS_2$ | 0.5 M TMACl | -0.8 to 0.0 V | 10 mVs$^{-1}$ | 11.0 Fg$^{-1}$ | [69] |
| $MoS_2$/$MoO_2$@CNT | 1 M KOH | 0.0 to 0.5 V | 0.5 Ag$^{-1}$ | 228.0 Fg$^{-1}$ | [70] |
| $MoO_3$/PPy supported $MoS_2$ | 2 M $H_2SO_4$ | -0.2 to 0.8 V | 5 mVs$^{-1}$ | 292.0 Fg$^{-1}$ | [71] |
| CC@$MoO_{3-x}$ | 3 M KCl | -1.0 to -0.3 V | 1 mVs$^{-1}$ | 7.1 Fg$^{-1}$ | [72] |
| $MoS_2$:$O@MoS_2$ | 1 M $Na_2SO_4$ | 0.0 – 0.9 V | 5 mVs$^{-1}$ | ~ 200.0 Fg$^{-1}$ | here |

## 4. Conclusion

The MoS$_2$:O@MoS$_2$ hybrids with different mass ratios were easily prepared, and their electrochemical performances were investigated in a three-electrode system in 1 M Na$_2$SO$_4$ electrolyte. The highest specific capacitance value of ~205 Fg$^{-1}$ was obtained for the MoS$_2$-O@MoS$_2$ (2:1) hybrid at a scan rate of 5 mVs$^{-1}$. Its spongy-like structure not only increased the effective surface area but also provided more active sites for charge storage. In addition, the hybrid maintained about 113% of its initial capacitance after 2000 charge-discharge cycles indicating excellent cyclic stability. Furthermore, MoS$_2$-O@MoS$_2$ (2:1) hybrid displayed an enhanced photoelectrochemical performance compared with the individual MoS$_2$ under light irradiation. The lower recombination rate of photogenerated carriers and efficient charge separation are responsible for this superiority. Thus, the introduced hybrid with improved (photo) electrochemical performance can be promising for energy storage and conversion applications.

## References

[1] B.C. O'Neill, T.R. Carter, K. Ebi, P.A. Harrison, E. Kemp-Benedict, K. Kok, E. Kriegler, B.L. Preston, K. Riahi, J. Sillmann, B.J. van Ruijven, D. van Vuuren, D. Carlisle, C. Conde, J. Fuglestvedt, C. Green, T. Hasegawa, J. Leininger, S. Monteith, R. Pichs-Madruga, Achievements and needs for the climate change scenario framework, Nature Climate Change, 10 (2020) 1074-1084, 10.1038/s41558-020-00952-0.
[2] D.P. Chatterjee, A.K. Nandi, A review on the recent advances in hybrid supercapacitors, Journal of Materials Chemistry A, 9 (2021) 15880-15918, 10.1039/D1TA02505H.
[3] S. Huang, Z. Wang, Y. Von Lim, Y. Wang, Y. Li, D. Zhang, H.Y. Yang, Recent Advances in Heterostructure Engineering for Lithium–Sulfur Batteries, Advanced Energy Materials, 11 (2021) 2003689, https://doi.org/10.1002/aenm.202003689.
[4] K. Dey, B. Roose, S.D. Stranks, Optoelectronic Properties of Low-Bandgap Halide Perovskites for Solar Cell Applications, Advanced Materials, 33 (2021) 2102300, https://doi.org/10.1002/adma.202102300.
[5] J.H. Kim, D. Hansora, P. Sharma, J.-W. Jang, J.S. Lee, Toward practical solar hydrogen production – an artificial photosynthetic leaf-to-farm challenge, Chemical Society Reviews, 48 (2019) 1908-1971, 10.1039/C8CS00699G.
[6] F. Ghasemi, M. Jalali, A. Abdollahi, S. Mohammadi, Z. Sanaee, S. Mohajerzadeh, A high performance supercapacitor based on decoration of MoS2/reduced graphene oxide with NiO nanoparticles, RSC Advances, 7 (2017) 52772-52781, 10.1039/C7RA09060A.
[7] E.E. Miller, Y. Hua, F.H. Tezel, Materials for energy storage: Review of electrode materials and methods of increasing capacitance for supercapacitors, Journal of Energy Storage, 20 (2018) 30-40, https://doi.org/10.1016/j.est.2018.08.009.

[8] Y. Huang, B. Wang, F. Liu, H. Liu, S. Wang, Q. Li, J. Cheng, L. Zhang, Fabrication of Rambutan-like Activated Carbon Sphere/Carbon Nanotubes and Their Application as Supercapacitors, Energy & Fuels, 35 (2021) 8313-8320, 10.1021/acs.energyfuels.1c00189.
[9] Y. Lei, X. Liang, L. Yang, P. Jiang, Z. Lei, S. Wu, J. Feng, Novel hierarchical porous carbon prepared by a one-step template route for electric double layer capacitors and Li–Se battery devices, Journal of Materials Chemistry A, 8 (2020) 4376-4385, 10.1039/C9TA13753J.
[10] K. Yan, X. Sun, S. Ying, W. Cheng, Y. Deng, Z. Ma, Y. Zhao, X. Wang, L. Pan, Y. Shi, Ultrafast microwave synthesis of rambutan-like CMK-3/carbon nanotubes nanocomposites for high-performance supercapacitor electrode materials, Scientific Reports, 10 (2020) 6227, 10.1038/s41598-020-63204-3.
[11] J. Zeng, C. Xu, T. Gao, X. Jiang, X.-B. Wang, Porous monoliths of 3D graphene for electric double-layer supercapacitors, Carbon Energy, 3 (2021) 193-224, https://doi.org/10.1002/cey2.107.
[12] X. Han, Q. Meng, X. Wan, B. Sun, Y. Zhang, B. Shen, J. Gao, Y. Ma, P. Zuo, S. Lou, G. Yin, Intercalation pseudocapacitive electrochemistry of Nb-based oxides for fast charging of lithium-ion batteries, Nano Energy, 81 (2021) 105635, https://doi.org/10.1016/j.nanoen.2020.105635.
[13] W. Wu, C. Wang, C. Zhao, L. Wang, J. Zhu, Y. Xu, Rational design of hierarchical FeCo2O4 nanosheets@NiO nanowhiskers core-shell heterostructure as binder-free electrodes for efficient pseudocapacitors, Electrochimica Acta, 370 (2021) 137789, https://doi.org/10.1016/j.electacta.2021.137789.
[14] A. Qian, Y. Pang, G. Wang, Y. Hao, Y. Liu, H. Shi, C.-H. Chung, Z. Du, F. Cheng, Pseudocapacitive Charge Storage in MXene–V2O5 for Asymmetric Flexible Energy Storage Devices, ACS Applied Materials & Interfaces, 12 (2020) 54791-54797, 10.1021/acsami.0c16959.
[15] A. Abdollahi, A. Abnavi, F. Ghasemi, S. Ghasemi, Z. Sanaee, S. Mohajerzadeh, Facile synthesis and simulation of MnO2 nanoflakes on vertically aligned carbon nanotubes, as a high-performance electrode for Li-ion battery and supercapacitor, Electrochimica Acta, 390 (2021) 138826, https://doi.org/10.1016/j.electacta.2021.138826.
[16] Y. Niu, H. Su, X. Li, J. Li, Y. Qi, Synthesis of porous α-MoO3 microspheres as electrode materials for supercapacitors, Journal of Alloys and Compounds, 898 (2022) 162863, https://doi.org/10.1016/j.jallcom.2021.162863.
[17] H.R. Barai, N.S. Lopa, F. Ahmed, N.A. Khan, S.A. Ansari, S.W. Joo, M.M. Rahman, Synthesis of Cu-Doped Mn3O4@Mn-Doped CuO Nanostructured Electrode Materials by a Solution Process for High-Performance Electrochemical Pseudocapacitors, ACS Omega, 5 (2020) 22356-22366, 10.1021/acsomega.0c02740.
[18] Lichchhavi, H. Lee, Y. Ohshita, A.K. Singh, P.M. Shirage, Transformation of Battery to High Performance Pseudocapacitor by the Hybridization of W18O49 with RuO2 Nanostructures, Langmuir, 37 (2021) 1141-1151, 10.1021/acs.langmuir.0c03056.
[19] X. Li, L. Zhao, J. Yu, X. Liu, X. Zhang, H. Liu, W. Zhou, Water Splitting: From Electrode to Green Energy System, Nano-Micro Letters, 12 (2020) 131, 10.1007/s40820-020-00469-3.
[20] F. Ghasemi, M. Hassanpour Amiri, Facile in situ fabrication of rGO/MoS2 heterostructure decorated with gold nanoparticles with enhanced photoelectrochemical performance, Applied Surface Science, 570 (2021) 151228, https://doi.org/10.1016/j.apsusc.2021.151228.
[21] Z. Masoumi, M. Tayebi, M. Kolaei, A. Tayyebi, H. Ryu, J.I. Jang, B.-K. Lee, Simultaneous Enhancement of Charge Separation and Hole Transportation in a W:α-Fe2O3/MoS2 Photoanode: A Collaborative Approach of MoS2 as a Heterojunction and W as a Metal Dopant, ACS Applied Materials & Interfaces, 13 (2021) 39215-39229, 10.1021/acsami.1c08139.
[22] B. Zhao, C. Feng, X. Huang, Y. Ding, Y. Bi, Coupling NiCo catalysts with carbon quantum dots on hematite photoanodes for highly efficient oxygen evolution, Journal of Materials Chemistry A, (2022), 10.1039/D1TA10039D.
[23] Y. Wang, B. Ren, J. Zhen Ou, K. Xu, C. Yang, Y. Li, H. Zhang, Engineering two-dimensional metal oxides and chalcogenides for enhanced electro- and photocatalysis, Science Bulletin, 66 (2021) 1228-1252, https://doi.org/10.1016/j.scib.2021.02.007.

[24] C.Y. Chot, M.N. Chong, A.K. Soh, K.W. Tan, J.D. Ocon, C. Saint, Facile synthesis and characterisation of functional MoO3 photoanode with self-photorechargeability, Journal of Alloys and Compounds, 838 (2020) 155624, https://doi.org/10.1016/j.jallcom.2020.155624.
[25] Y. Zhu, Y. Yao, Z. Luo, C. Pan, J. Yang, Y. Fang, H. Deng, C. Liu, Q. Tan, F. Liu, Y. Guo, Nanostructured MoO3 for Efficient Energy and Environmental Catalysis, Molecules, 25 (2020) 18,
[26] O. Samy, A. El Moutaouakil, A Review on MoS2 Energy Applications: Recent Developments and Challenges, Energies, 14 (2021) 4586,
[27] D.B. Sulas-Kern, E.M. Miller, J.L. Blackburn, Photoinduced charge transfer in transition metal dichalcogenide heterojunctions – towards next generation energy technologies, Energy & Environmental Science, 13 (2020) 2684-2740, 10.1039/D0EE01370F.
[28] F. Ghasemi, S. Mohajerzadeh, Sequential Solvent Exchange Method for Controlled Exfoliation of MoS2 Suitable for Phototransistor Fabrication, ACS Applied Materials & Interfaces, 8 (2016) 31179-31191, 10.1021/acsami.6b07211.
[29] X. Zou, J. Zhang, X. Zhao, Z. Zhang, MoS2/RGO composites for photocatalytic degradation of ranitidine and elimination of NDMA formation potential under visible light, Chemical Engineering Journal, 383 (2020) 123084, https://doi.org/10.1016/j.cej.2019.123084.
[30] D. Cao, Q. Wang, S. Zhu, X. Zhang, Y. Li, Y. Cui, Z. Xue, S. Gao, Hydrothermal construction of flower-like MoS2 on TiO2 NTs for highly efficient environmental remediation and photocatalytic hydrogen evolution, Separation and Purification Technology, 265 (2021) 118463, https://doi.org/10.1016/j.seppur.2021.118463.
[31] F. Han, Z. Song, M.H. Nawaz, M. Dai, D. Han, L. Han, Y. Fan, J. Xu, D. Han, L. Niu, MoS2/ZnO-Heterostructures-Based Label-Free, Visible-Light-Excited Photoelectrochemical Sensor for Sensitive and Selective Determination of Synthetic Antioxidant Propyl Gallate, Analytical Chemistry, 91 (2019) 10657-10662, 10.1021/acs.analchem.9b01889.
[32] H. Tran Huu, M.D.N. Thi, V.P. Nguyen, L.N. Thi, T.T.T. Phan, Q.D. Hoang, H.H. Luc, S.J. Kim, V. Vo, One-pot synthesis of S-scheme MoS2/g-C3N4 heterojunction as effective visible light photocatalyst, Scientific Reports, 11 (2021) 14787, 10.1038/s41598-021-94129-0.
[33] T. Nam Trung, F.Z. Kamand, T.M. Al tahtamouni, Elucidating the mechanism for the chemical vapor deposition growth of vertical MoO2/MoS2 flakes toward photoelectrochemical applications, Applied Surface Science, 505 (2020) 144551, https://doi.org/10.1016/j.apsusc.2019.144551.
[34] Y. Wang, Y. Xie, Electroactive FeS2-modified MoS2 nanosheet for high-performance supercapacitor, Journal of Alloys and Compounds, 824 (2020) 153936, https://doi.org/10.1016/j.jallcom.2020.153936.
[35] D. Sarkar, D. Das, S. Das, A. Kumar, S. Patil, K.K. Nanda, D.D. Sarma, A. Shukla, Expanding Interlayer Spacing in MoS2 for Realizing an Advanced Supercapacitor, ACS Energy Letters, 4 (2019) 1602-1609, 10.1021/acsenergylett.9b00983.
[36] Y. Zhang, P. Chen, Q. Wang, Q. Wang, K. Zhu, K. Ye, G. Wang, D. Cao, J. Yan, Q. Zhang, High-Capacity and Kinetically Accelerated Lithium Storage in MoO3 Enabled by Oxygen Vacancies and Heterostructure, Advanced Energy Materials, 11 (2021) 2101712, https://doi.org/10.1002/aenm.202101712.
[37] S.V.P. Vattikuti, P.C. Nagajyothi, P. Anil Kumar Reddy, M. Kotesh Kumar, J. Shim, C. Byon, Tiny MoO3 nanocrystals self-assembled on folded molybdenum disulfide nanosheets via a hydrothermal method for supercapacitor, Materials Research Letters, 6 (2018) 432-441, 10.1080/21663831.2018.1477848.
[38] X. Liu, Y. Wu, H. Wang, Y. Wang, C. Huang, L. Liu, Z. Wang, Two-dimensional β-MoO3@C nanosheets as high-performance negative materials for supercapacitors with excellent cycling stability, RSC Advances, 10 (2020) 17497-17505, 10.1039/D0RA01258K.
[39] Y. Tian, X. Yang, A. Nautiyal, Y. Zheng, Q. Guo, J. Luo, X. Zhang, One-step microwave synthesis of MoS2/MoO3@graphite nanocomposite as an excellent electrode material for supercapacitors, Advanced Composites and Hybrid Materials, 2 (2019) 151-161, 10.1007/s42114-019-0075-4.
[40] C.D. Quilty, L.M. Housel, D.C. Bock, M.R. Dunkin, L. Wang, D.M. Lutz, A. Abraham, A.M. Bruck, E.S. Takeuchi, K.J. Takeuchi, A.C. Marschilok, Ex Situ and Operando XRD and XAS Analysis of MoS2:

A Lithiation Study of Bulk and Nanosheet Materials, ACS Applied Energy Materials, 2 (2019) 7635-7646, 10.1021/acsaem.9b01538.
[41] Z. Tang, X. Deng, Y. Zhang, X. Guo, J. Yang, C. Zhu, J. fan, Y. Shi, B. Qing, F. Fan, MoO3 nanoflakes coupled reduced graphene oxide with enhanced ethanol sensing performance and mechanism, Sensors and Actuators B: Chemical, 297 (2019) 126730, https://doi.org/10.1016/j.snb.2019.126730.
[42] Q. Wang, C. Zhou, X.H. Yan, J.J. Wang, D.F. Wang, X.X. Yuan, X.N. Cheng, TiO2 Nanoparticles Modified MoO3 Nanobelts as Electrode Materials with Superior Performances for Supercapacitors, Energy Technology, 6 (2018) 2367-2373, https://doi.org/10.1002/ente.201800346.
[43] A. Ohtake, T. Mano, Y. Sakuma, Strain relaxation in InAs heteroepitaxy on lattice-mismatched substrates, Scientific Reports, 10 (2020) 4606, 10.1038/s41598-020-61527-9.
[44] C. Zhang, S. Wu, L. Tao, G.M. Arumugam, C. Liu, Z. Wang, S. Zhu, Y. Yang, J. Lin, X. Liu, R.E.I. Schropp, Y. Mai, Fabrication Strategy for Efficient 2D/3D Perovskite Solar Cells Enabled by Diffusion Passivation and Strain Compensation, Advanced Energy Materials, 10 (2020) 2002004, https://doi.org/10.1002/aenm.202002004.
[45] F. Ghasemi, Vertically aligned carbon nanotubes, MoS2–rGo based optoelectronic hybrids for NO2 gas sensing, Scientific Reports, 10 (2020) 11306, 10.1038/s41598-020-68388-2.
[46] X. Guan, Y. Ren, S. Chen, J. Yan, G. Wang, H. Zhao, W. Zhao, Z. Zhang, Z. Deng, Y. Zhang, Y. Dai, L. Zou, R. Chen, C. Liu, Charge separation and strong adsorption-enhanced MoO3 visible light photocatalytic performance, Journal of Materials Science, 55 (2020) 5808-5822, 10.1007/s10853-020-04418-8.
[47] F. Ghasemi, A. Abdollahi, S. Mohajerzadeh, Controlled Plasma Thinning of Bulk MoS2 Flakes for Photodetector Fabrication, ACS Omega, 4 (2019) 19693-19704, 10.1021/acsomega.9b02367.
[48] F. Ghasemi, A. Abdollahi, A. Abnavi, S. Mohajerzadeh, Y. Abdi, Ultrahigh Sensitive MoS$_2$/rGo Photodetector Based on Aligned CNT Contacts, IEEE Electron Device Letters, 39 (2018) 1465-1468, 10.1109/LED.2018.2857844.
[49] Z. Pan, J.A. Röhr, Z. Ye, Z.S. Fishman, Q. Zhu, X. Shen, S. Hu, Elucidating charge separation in particulate photocatalysts using nearly intrinsic semiconductors with small asymmetric band bending, Sustainable Energy & Fuels, 3 (2019) 850-864, 10.1039/C9SE00036D.
[50] D. Mandal, P. Routh, A.K. Mahato, A.K. Nandi, Electrochemically modified graphite paper as an advanced electrode substrate for supercapacitor application, Journal of Materials Chemistry A, 7 (2019) 17547-17560, 10.1039/C9TA04496E.
[51] A. Merazga, J. Al-Zahrani, A. Al-Baradi, B. Omer, A. Badawi, S. Al-Omairy, Optical band-gap of reduced graphene oxide/TiO2 composite and performance of associated dye-sensitized solar cells, Materials Science and Engineering: B, 259 (2020) 114581, https://doi.org/10.1016/j.mseb.2020.114581.
[52] K. Shen, J. Ding, S. Yang, 3D Printing Quasi-Solid-State Asymmetric Micro-Supercapacitors with Ultrahigh Areal Energy Density, Advanced Energy Materials, 8 (2018) 1800408, https://doi.org/10.1002/aenm.201800408.
[53] X. Chang, W. Li, Y. Liu, M. He, X. Zheng, J. Bai, Z. Ren, Hierarchical NiCo2S4@NiCoP core-shell nanocolumn arrays on nickel foam as a binder-free supercapacitor electrode with enhanced electrochemical performance, Journal of Colloid and Interface Science, 538 (2019) 34-44, https://doi.org/10.1016/j.jcis.2018.11.080.
[54] P. Hota, M. Miah, S. Bose, D. Dinda, U.K. Ghorai, Y.-K. Su, S.K. Saha, Ultra-small amorphous MoS2 decorated reduced graphene oxide for supercapacitor application, Journal of Materials Science & Technology, 40 (2020) 196-203, https://doi.org/10.1016/j.jmst.2019.08.032.
[55] S. Sun, J. Luo, Y. Qian, Y. Jin, Y. Liu, Y. Qiu, X. Li, C. Fang, J. Han, Y. Huang, Metal–Organic Framework Derived Honeycomb Co9S8@C Composites for High-Performance Supercapacitors, Advanced Energy Materials, 8 (2018) 1801080, https://doi.org/10.1002/aenm.201801080.
[56] S. Huo, Y. Zhao, M. Zong, B. Liang, X. Zhang, I.U. Khan, X. Song, K. Li, Boosting supercapacitor and capacitive deionization performance of hierarchically porous carbon by polar surface and structural engineering, Journal of Materials Chemistry A, 8 (2020) 2505-2517, 10.1039/C9TA12170F.

[57] R. Rajagopal, K.-S. Ryu, Morphologically engineered cactus-like MnO2 nanostructure as a high-performance electrode material for energy-storage applications, Journal of Energy Storage, 32 (2020) 101880, https://doi.org/10.1016/j.est.2020.101880.

[58] B.-A. Mei, O. Munteshari, J. Lau, B. Dunn, L. Pilon, Physical Interpretations of Nyquist Plots for EDLC Electrodes and Devices, The Journal of Physical Chemistry C, 122 (2018) 194-206, 10.1021/acs.jpcc.7b10582.

[59] M.R. Pallavolu, N. Gaddam, A.N. Banerjee, R.R. Nallapureddy, S.W. Joo, Superior energy-power performance of N-doped carbon nano-onions-based asymmetric and symmetric supercapacitor devices, International Journal of Energy Research, 46 (2022) 1234-1249, https://doi.org/10.1002/er.7242.

[60] A.U. Ammar, I.D. Yildirim, F. Bakan, E. Erdem, ZnO and MXenes as electrode materials for supercapacitor devices, Beilstein Journal of Nanotechnology, 12 (2021) 49-57,

[61] M. Buldu-Aktürk, Ö. Balcı-Çağıran, E. Erdem, EPR investigation of point defects in HfB2 and their roles in supercapacitor device performances, Applied Physics Letters, 120 (2022) 153901, 10.1063/5.0089931.

[62] M. Buldu-Aktürk, M. Toufani, A. Tufani, E. Erdem, ZnO and reduced graphene oxide electrodes for all-in-one supercapacitor devices, Nanoscale, 14 (2022) 3269-3278, 10.1039/D2NR00018K.

[63] S. Kasap, I.I. Kaya, S. Repp, E. Erdem, Superbat: battery-like supercapacitor utilized by graphene foam and zinc oxide (ZnO) electrodes induced by structural defects, Nanoscale Advances, 1 (2019) 2586-2597, 10.1039/C9NA00199A.

[64] M. Toufani, S. Kasap, A. Tufani, F. Bakan, S. Weber, E. Erdem, Synergy of nano-ZnO and 3D-graphene foam electrodes for asymmetric supercapacitor devices, Nanoscale, 12 (2020) 12790-12800, 10.1039/D0NR02028A.

[65] D. Kesavan, V.K. Mariappan, P. Pazhamalai, K. Krishnamoorthy, S.-J. Kim, Topochemically synthesized MoS2 nanosheets: A high performance electrode for wide-temperature tolerant aqueous supercapacitors, Journal of Colloid and Interface Science, 584 (2021) 714-722, https://doi.org/10.1016/j.jcis.2020.09.088.

[66] R.B. Choudhary, M. Majumder, A.K. Thakur, Two-Dimensional Exfoliated MoS2 Flakes Integrated with Polyindole for Supercapacitor Application, ChemistrySelect, 4 (2019) 6906-6912, https://doi.org/10.1002/slct.201901558.

[67] P. Pazhamalai, K. Krishnamoorthy, S. Manoharan, S.J. Kim, High energy symmetric supercapacitor based on mechanically delaminated few-layered MoS2 sheets in organic electrolyte, Journal of Alloys and Compounds, 771 (2019) 803-809, https://doi.org/10.1016/j.jallcom.2018.08.203.

[68] L. Wang, Y. Ma, M. Yang, Y. Qi, Titanium plate supported MoS2 nanosheet arrays for supercapacitor application, Applied Surface Science, 396 (2017) 1466-1471, https://doi.org/10.1016/j.apsusc.2016.11.193.

[69] P. Iamprasertkun, W. Hirunpinyopas, A.M. Tripathi, M.A. Bissett, R.A.W. Dryfe, Electrochemical intercalation of MoO3-MoS2 composite electrodes: Charge storage mechanism of non-hydrated cations, Electrochimica Acta, 307 (2019) 176-187, https://doi.org/10.1016/j.electacta.2019.03.141.

[70] Y. Tian, H. Du, M. Zhang, Y. Zheng, Q. Guo, H. Zhang, J. Luo, X. Zhang, Microwave synthesis of MoS2/MoO2@CNT nanocomposites with excellent cycling stability for supercapacitor electrodes, Journal of Materials Chemistry C, 7 (2019) 9545-9555, 10.1039/C9TC02391G.

[71] F.N. Indah Sari, J.-M. Ting, High performance asymmetric supercapacitor having novel 3D networked polypyrrole nanotube/N-doped graphene negative electrode and core-shelled MoO3/PPy supported MoS2 positive electrode, Electrochimica Acta, 320 (2019) 134533, https://doi.org/10.1016/j.electacta.2019.07.044.

[72] J.-C. Liu, H. Li, M. Batmunkh, X. Xiao, Y. Sun, Q. Zhao, X. Liu, Z.-H. Huang, T.-Y. Ma, Structural engineering to maintain the superior capacitance of molybdenum oxides at ultrahigh mass loadings, Journal of Materials Chemistry A, 7 (2019) 23941-23948, 10.1039/C9TA04835A.

Graphical Abstract

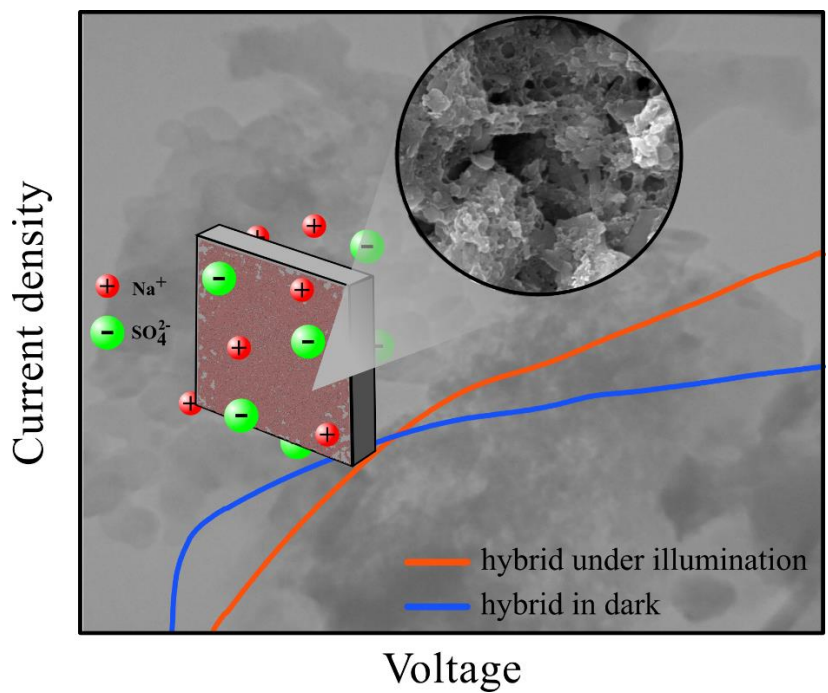